\documentclass[journal = jacsat]{achemso} 

\usepackage[T1]{fontenc} 
\usepackage{xcolor}
\usepackage{textcomp}
\usepackage{tabulary,booktabs}
\usepackage{array}
\usepackage{multirow}

\author{Faizan Khan}
\affiliation{Department of Chemical Engineering, Indian Institute of Technology Ropar, Rupnagar-140001, Punjab, India}
\author{Chandra Shekhar}
\affiliation{Department of Chemical Engineering, Indian Institute of Technology Ropar, Rupnagar-140001, Punjab, India}
\author{Tarak Mondal}
\affiliation{Department of Chemical Engineering, Indian Institute of Technology Ropar, Rupnagar-140001, Punjab, India}
\author{Manigandan Sabapathy}
\affiliation{Department of Chemical Engineering, Indian Institute of Technology Ropar, Rupnagar-140001, Punjab, India}
\email{mani@iitrpr.ac.in}

\title {Removal of industrial dye and pharmaceutical product using the nano and micron-sized PS rough particles studded with Pt nanoparticles}
\keywords{Polystyrene, iron oxide, platinum nanoparticles, magnetically-responsive rough particles, methylene blue, tetracycline}

\begin{document}

\begin{sloppypar}


\begin{abstract}
We show that the rough particles studded with platinum nanoparticles can be fabricated straightforwardly and in a single step at room temperature. These rough particles displayed a good catalytic power (100\% removal efficiency) against a model industrial dye (methylene blue) and pharmaceutical residue (tetracycline) within a reasonable time scale. Further, we illustrate the effects of particle size, concentration, and contact patterns on the performance of rough catalytic particles. The semi-batch conditions favoured the complete decomposition of tetracycline within 40 min, but the batch-wise operation offered a good contacting pattern for methylene blue yielding a maximal output within 10 min. The kinetics of the heterogeneous catalytic process modelled by Langmuir-Hinshelwood kinetics predicts that the given methylene blue decomposition reaction induced by the rough particles follows the pseudo-first-order kinetics. The rate constants for the reaction catalyzed by 0.6 and 1.0 $\mu$m-sized rough particles are 0.048 and 0.032 min$^{-1}$, respectively. Furthermore, we established the proof-of-concept using magnetically-responsive rough particles for real-time applications, including decontamination and recovery of catalyst particles via an externally applied magnetic field in one cycle. Our proposed method helps achieve a near-100\% degrading efficiency within 10 to 40 min at minimal catalytic particle concentration, i.e., 200 ppm. Since we can turn the rough particles into super-paramagnetic, we can recover and reuse them for several wastewater treatment cycles without incurring any running costs.
\end{abstract}

\section{Introduction}

The rapid rise of contamination in the groundwater due to industrial dyes and pharmaceutical products generated from textile and pharma industries has been a most pressing issue globally that it has urged the intervention of policymakers, researchers, and engineers around the globe\cite{visa2014hydrothermally,liu2021treatment}. Although the contamination of several azo-based dyes and pharmaceutical residues is supposed to be a matter of concern, the primary issue that remains to be tackled is establishing efficient removal of the contaminants from the wastewater as it poses a severe threat to ecology. In recent years, much attention has been given to the advanced oxidation processes (AOPs)\cite{minero2005fe} over traditional water treatment methods such as mechanical separation, coagulation, and filtration. The various techniques of AOPs include 1) photolysis\cite{peter2017uv,mohammed2021photolysis}, 2) Fenton reaction\cite{hsieh2012fept,saleh2019degradation}, 3) photo-Fenton\cite{saleh2019degradation,kirchon2020effect}, 4) photo-catalysis\cite{peter2017uv,ikhlaqa2020combined}, 5) ozonation\cite{adelin2020ozonation,ikhlaqa2020combined}, and 6) sonochemistry\cite{saleh2019degradation,qian2019sonochemical}. Despite the advancement in these techniques being well received under AOPs, each method has merits and demerits. For instance, high energy demand, strict pH range (Fenton)\cite{liu2016degradation}, high H$_{2}$O$_{2}$ consumption (Fenton/photo-Fenton), and cost of implementation.

Among several AOPs, Fenton reaction-based methods are found to be the most efficient one for the removal of organic pollutants\cite{pignatello2006advanced}. Nevertheless, this method is not much attractive due to a few shortcomings such as 1) tendency to aggregate and sediment against gravity, 2) process condition that demands narrow pH range of 2-3 to be maintained, and 3) high peroxide consumptions. On the contrary, heterogeneous Fenton-like catalyst-assisted reaction based on iron oxides\cite{pignatello2006advanced}, transition metal oxides\cite{GIORDANO2007240}, and the alloy made up of zero-charge metal/metal-iron oxides \cite{ai2007fe} are promising candidates owing to its increased electrocatalytic activity even at higher pH. Toda et al. reported the enhanced electrocatalytic activity of Pt alloyed with Fe electrodes. Based on the analysis by these authors, the enhanced electrocatalytic activity is due to increased d-electron vacancy of the Pt surface induced by alloying the metal with Fe\cite{Toda_1999}. Furthermore, the efficiency of the Fenton reaction (based on Fe$_{3}$O$_{4}$ nanoparticles) is expected to decrease as the considerable amount of hydroxyl and perhydroxyl radicals formed during reaction are involved in converting Fe$^{3+}$ back to Fe$^{2+}$, in addition to organic molecules. However, the decomposition reaction catalyzed by platinum-x hybrid particles instead of Fenton, eliminate the perhydroxyl formation and the unwanted reaction step\cite{hsieh2012fept}. In 2012, \citeauthor {hsieh2012fept} reported the electrocatalytic activity of Fe-Pt nanoparticles (Fenton-like) for the first time with a maximum efficiency of 90\% by the end of the 90 min reaction time\cite{hsieh2012fept}. The study of the decomposition of MB based on the Pt/Fe route was not pursued by many researchers considering the cost of synthesis involved, despite knowing the fact that the issue associated with the cost of production could be eliminated by employing a magnetically responsive core-shell particle system decorated with platinum nanoparticles, as the particles can be recovered by applying magnetic force externally. Such particles are regarded as rough particles with controlled surface deposition. We aim to demonstrate the chemical decomposition of MB (an example system of azo dyes) and tetracycline (an example system of pharmaceutical drugs) using nano and micron-sized polystyrene (PS) rough particles decorated with platinum as zero charge metal catalyst particles. Since the organics mentioned above is chemically deactivated, there is no need to regenerate the catalyst particles once the reaction is completed. Furthermore, since the particles can be made magnetically responsive and easily separated by applying magnetic force externally, it is possible to carry out the treatment process continuously with less or no manual intervention without regeneration. Thus, as outlined above, the method of heterogeneous catalyst particles is a cost-effective alternative, as the costs spent are predominantly linked to the capital budget or one-time investment. 

In the context of the synthesis of rough particles concerning immobilization of metallic nanoparticles, a common approach has been to decorate the surface of core particles with gold (Au) or any other metallic nanoparticles via electrostatic interaction or covalent bonding\cite{dokoutchaev1999colloidal,lee2009facile,bao2014facile}. These routes demand the core particles be cationic to adsorb/deposit the gold nanoparticles (anionic) to make suitable rough colloids. However, it is of our interest to immobilize platinum (Pt) instead of Au nanoparticles to explore its application in the field of contaminant removal. The synthesis route follows the covalent bonding route in which the Pt nanoparticles are deposited by the chemical reduction of platinic acid using a reducing agent such as sodium borohydride (NaBH$_{4}$). In 2012, \citeauthor{hsieh2012fept} devised Fe-Pt nanoparticles to deactivate MB dye to achieve greater reaction extent\cite{hsieh2012fept}. However, the method offered by \citeauthor{hsieh2012fept} is slightly complicated and involves heating up to 297$^{\circ}$C. Therefore, it is essential to study the heterogeneous catalysts of many kinds to find the suitability of the method to tackle various pollutants. Thus, we have chosen to extend the studies beyond Fe-based supports to get insight into the degradation kinetics against other organics removal. For instance, the as-synthesized Pt-based rough particles have shown good catalytic activity in decomposing hydrogen peroxide, which helps degrade methylene blue (MB) and tetracycline (TC). Our suggested technique achieves a near-100\% degrading efficiency within 10 to 40 minutes at lower concentrations (200 ppm), making it a viable alternative to Fenton-based AOPs.     

The remainder of the manuscript is structured as follows: To provide the reader a better understanding, we first address the synthesis, characterization, and application of Pt-decorated PS rough particles for wastewater treatment. Consequently, we show the proof-of-concept (POC) by using magnetically-responsive rough particles (MR-RP) for the application. In addition to the application of discolouration or deactivation, this POC portion addresses the separation of catalyst particles under an externally applied magnetic force.
\section{Materials and Methods}
\subsection{Materials}
We used Invitrogen\texttrademark{} grade amidine functionalized polystyrene (PS) latex particles purchased from Thermo Fisher Scientific, India, to prepare catalytic-enabled rough particles for treating the wastewater containing specific contaminants. We used polystyrene particles of different sizes for carrying out the decomposition reaction. The size of the particles employed, as per the manufacturer's specifications, are 1.0, 0.5, and 0.02 $\mu$m. However, the size of the particles determined based on image analysis using a scanning electron microscope (SEM) and dynamic light scattering (DLS) technique are 1.0 $\pm$ 0.1, 0.579 $\pm$ 0.006, and 0.022 $\pm$ 0.001 $\mu$m. An electrophoretic study was carried out to determine the particles' potential at the boundary between the shear and diffuse plane, i.e., Zeta potential ($\zeta$). The $\zeta$ potential of the particles determined for PS particles of size 1.0, 0.6 (rounded-off) and 0.02 $\mu$m in 1 mM NaCl electrolyte medium were 46 $\pm$ 3, 55.8 $\pm$ 3.2 and 42.7 $\pm$ 1 mV, respectively. To prepare a core-shell assembly of magnetically-responsive rough particles, we employed iron oxide (IO) nanoparticles purchased from Sigma Aldrich Chemicals Pvt. Ltd., India. Polydiallyldimethylammonium chloride (DADMAC), gifted by Dr. Sarang Gumfekar, IIT Ropar, was used to modify the IO particles by imparting positive sites to the surface. Chloroplatinic acid hexahydrate (H$_{2}$PtCl$_{6}$.6H$_{2}$O), used as precursor solution to deposit platinum (Pt) nanoparticles on the PS surface, was procured from Sigma-Aldrich Chemicals Pvt. Ltd., India. Sodium borohydride (NaBH$_{4}$), obtained from Sigma-Aldrich Chemicals Pvt. Ltd., was used as a reducing agent. Hydrogen peroxide (30 w/v \%), used in the decomposition reaction mixture to generate hydroxy radical, was procured from Sigma-Aldrich Chemicals Pvt. Ltd., India. The model pollutants used to demonstrate the deactivation kinetics, such as methylene blue and tetracycline, were procured from Sigma-Aldrich Chemicals Pvt. Ltd., India. A hydrogen peroxide test strip (MQuant\textsuperscript{\textregistered} Peroxide-Test) received from Sigma-Aldrich Chemicals Pvt. Ltd., India, was used to quantitatively predict the catalytic-assisted decomposition kinetics under the presence of rough colloids. All reagents received were analytical grades and used without any further purification. We used deionized water obtained from a Smart2Pure\texttrademark{} water purification system (Make: Thermo Fisher Scientific, Model: Smart2Pure 12 UV/UF) for all experiments.

\subsection{Synthesis of PS rough particles decorated with Pt}
First, we prepared a chloroplatinic acid hexahydrate solution using DI water such that the precursor concentration in the stock solution was 10 mM. We devised a wet-chemical method to achieve a uniform deposition of Pt nanoparticles on the PS surface in an endless array of monolayer fashion. The deposition of Pt nanoparticles was realized by using the method reported by our group that involves the deposition of gold (Au) nanoparticles\cite{sabapathy2015visualization}. We used a slightly different strategy, including a wet-chemical deposition by nucleating Pt nanoparticles instead of Au on the PS surface. Since both the precursors of Au and Pt bear a net negative charge, favouring the attachment of ions over the positive sites of PS due to net electrostatic attraction, the modified protocol helped us synthesize Pt-decorated PS particles in a single step. Unlike the work of \citeauthor{lu2003colloidal}, our modified protocol is devoid of pre-deposition of Au nanoparticles as seeds to catalyze the reduction reaction of PtCl$_{6}^{2-}$ ions. Scheme \ref{scheme} depicts the schematic description of making rough particles and their application towards wastewater treatment. As explained in Scheme \ref{scheme}A, a known concentration of platinum precursor (0.5 mM), PS particles (0.5 mg/mL) are mixed using a magnetic stirrer for up to 30 min before adding a reducing agent, sodium borohydride, to the mixture. Subsequently, after the equilibration of the mixture for about 30 min, the chemical reduction reaction is induced by adding 5 mL of 10 mM sodium borohydride solution completely into the beaker containing the reaction mixture. The reaction is allowed to continue until the solution becomes clear. To ensure the reaction is complete, we have used sodium borohydride ten times in excess. Additionally, we have given sufficient time for reaction, i.e., 4 hr. We have maintained the reaction time of four hours in all the experiments that involve the deposition of Pt on the PS surface. Note: The reaction mixture becomes transparent and clear when the chemical reduction process continues for up to four hours. Further, we allowed the samples to centrifugation to separate the modified PS particles by removing the supernatant and washing them several times. The particles recovered in this way are further utilized to study the kinetics of MB and TC decomposition. Table \ref{table} presents the size and zeta potential ($\zeta$) of the bare PS and the rough PS particles.

\subsection{Selection of target pollutants}

We chose to work with two representative pollutants from the family of azo dyes and antibiotic drugs, such as MB and TC, respectively, for the entire decomposition studies. We endeavoured to understand the role of rough particles in the chemical deactivation of these target pollutants. The initial pollutants concentration was maintained constant at 15 $\mu$M throughout. We studied the degradation of these pollutants in batch and semi-batch conditions to replicate the real-time process of large-scale operation. 

\subsection{Heterogeneous batch degradation process}

All degradation experiments that involve batch-wise operations were performed in different initial volumes, such as 25, 50 and 100 mL. The rough particle concentrations employed were 200, 100, and 50 ppm, respectively. For the studies, we prepare the required mixture using the rough particles at appropriate concentrations along with the target pollutants with known initial concentration, i.e. 15 $\mu$M, to induce the decomposition reaction in the presence of a known amount of H$_{2}$O$_{2}$, 5 vol\% per the total volume mentioned above. The reaction is set to proceed for the desired time by adding the radical generator, i.e., H$_{2}$O$_{2}$. We continuously stirred the reaction mixture using the stirrer at a programmed speed of 140 RPM. The entire reaction was conducted at room temperature, 25$^{\circ}$C, throughout the desired duration.

\subsection{Heterogeneous semi-batch degradation process}

All degradation experiments that involved semi-batch operations were performed in different initial dosing volume at a uniform interval of 10 min until completion. The rough particles concentration employed were 200, 100, and 50 PPM, respectively. For the studies, we prepare the required mixture using the rough particles at appropriate concentrations along with the target pollutants with known initial concentration, i.e. 15 $\mu$M, to induce the decomposition reaction in the presence of a known amount of H$_{2}$O$_{2}$, 5 vol\% per the total initial volume. The reaction is initiated to proceed for the desired length by adding the radical generator, i.e., H$_{2}$O$_{2}$ at a desired dosing volume. Using the stirrer, we stirred the reaction mixture nonstop at a programmed speed of 140 RPM. The entire decomposition reaction was conducted at room temperature, 25$^{\circ}$C, throughout the desired duration.

Scheme \ref{scheme}B and E depicts the schematic description showing the catalytic action of PS rough particles and the process of discoloration of MB due to contact with the modified PS particles of varying sizes, respectively. 

\subsection{Hydrogen peroxide decomposition}

To understand the kinetics of H$_{2}$O$_{2}$ decomposition, we decoupled the decomposition reactions of MB and TC by eliminating them in the lab-scale reactor. Subsequently, in a separate set of experimental studies, the reaction flasks containing H$_{2}$O$_{2}$ and rough particles were employed to conduct studies at different lengths of time. We recorded the concentration vs time of H$_{2}$O$_{2}$ as the reaction proceeds to understand the effect of contacting patterns on the catalyst-assisted H$_{2}$O$_{2}$ decomposition reaction. 

\subsection{Characterization}

The zeta potential of bare PS and rough PS particles were measured based on the electrophoretic light scattering (ELS) technique using Zetasizer procured from Malvern Instruments, Model: Zetasizer Nano ZSP. The average diameter of PS of size 0.6 $\mu$m and 1.0 $\mu$m were determined using a high-resolution scanning electron microscope (HRSEM), Make: FEI, USA and Model: Inspect F50. The hydrodynamic radius of PS of size 22 nm was deduced using the dynamic light scattering (DLS), Make: Malvern Instruments and Model: Zetasizer Nano ZSP. The kinetics of decomposition reactions of MB and TC were studied by capturing concentration as a function of decomposition times of MB and TC using UV-VIS spectrophotometers (Make: Hach company US, Model: DR3900 \& Make: PerkinElmer Inc., USA, Model: LAMBDA 365)  

\begin{scheme}
\centering
\includegraphics[width=\linewidth]{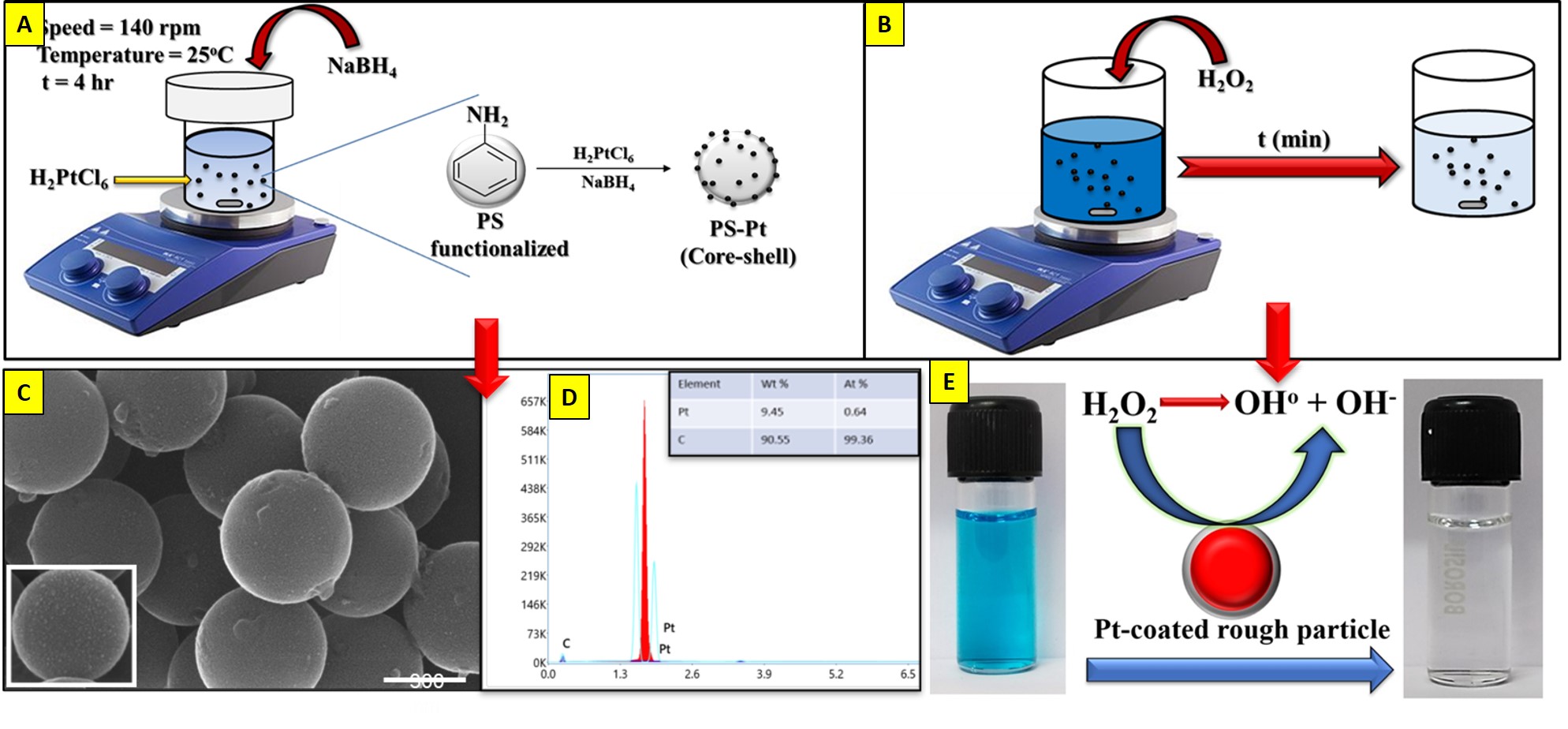}
\caption{Schematic description showing the process of synthesizing rough particles and their application towards deactivation of the target pollutants. A) Schematic illustration showing the synthesis methodology of making rough particles. B) Schematic illustration showing the application of rough particles in treating wastewater containing the target pollutants. C and D) FESEM and EDX spectrum of 1 $\mu$m rough particles modified with Pt nanoparticles, respectively. E) Vial pictures showing the discoloration of MB dye in the presence of rough particles.}
\label{scheme}
\end{scheme}

\begin{table}[htbp]
  \centering
  \scalebox{1.2}{
  \caption{Average size and zeta potential of the PS particles}
  \label{table}
    \begin{tabular}{cccc}
    \toprule
    \multirow{2}[4]{*}{Type} & \multirow{2}[4]{*}{Avg. Size ($\mu$m)} & \multicolumn{2}{c}{Zeta potential (mV)} \\
\cmidrule{3-4}          &       & Bare  & Modified \\
    \midrule
    \multirow{3}[6]{*}{PS} & 0.02 $\pm$ 0.009  & 42.7 $\pm$ 1.0 & -33 $\pm$ 0.7 \\
\cmidrule{2-4}          & 0.6 $\pm$ 0.006 & 55.8 $\pm$ 3.2 & -33 $\pm$ 0.3 \\
\cmidrule{2-4}          & 1.0 $\pm$ 0.1 & 46 $\pm$ 3    & -35 $\pm$ 0.4 \\
\midrule
IO & 0.06 $\pm$ 0.001  & -2.7 $\pm$ 1.8 & 33.2 $\pm$ 2.2 \\
    \bottomrule
    \end{tabular}}
\end{table}

\section{Results \& Discussion}

\begin{figure}
\centering
\includegraphics[width=\linewidth]{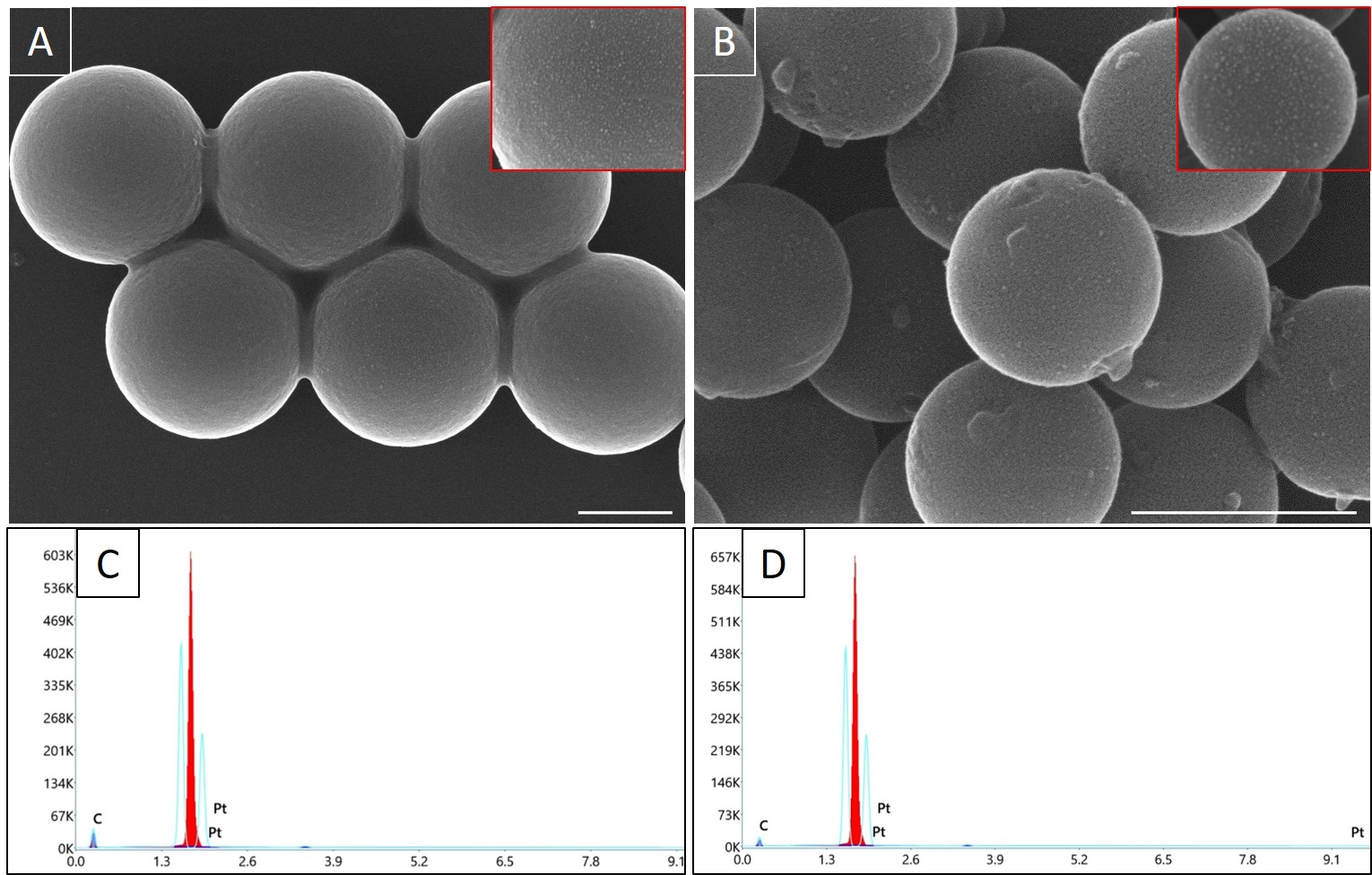}
\caption{Structural characterization of modified PS particles. A and B) FESEM images of 1.0 and 0.6 $\mu$m, respectively. C and D) EDX spectrum of 1.0 and 0.6 $\mu$m, respectively. Scale bars correspond to 500 nm.}
\label{SEM}
\end{figure}

FESEM was used to determine the size of PS particles and the morphological changes in PS rough particles due to the reaction. EDX was performed concurrently to identify the presence of Pt nanoparticles. Figure \ref{SEM} displays the surface morphology of surface-engineered PS rough particles decorated with Pt nanoparticles in the form of islands (Figures \ref{SEM}A and \ref{SEM}B) and the EDX spectrum used to validate the deposition of Pt nanoparticles on the surface of PS (Figures \ref{SEM}C and \ref{SEM}D). As depicted in Figures \ref{SEM}C and \ref{SEM}D, the distinctive peaks at 1.5 KeV indicate the existence of Pt nanoparticles.

\begin{figure}
\centering
\includegraphics[width=\linewidth]{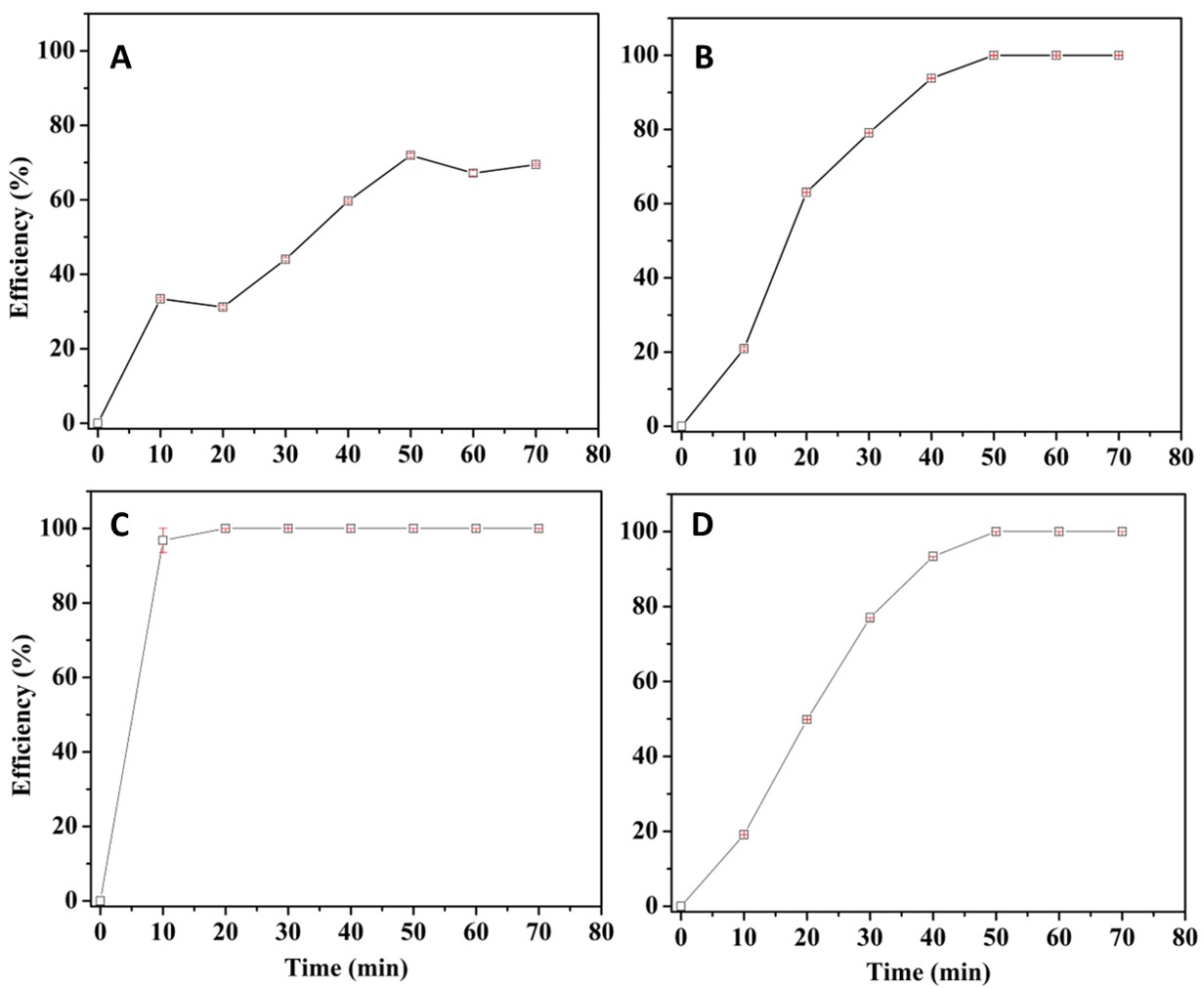}
\caption{Decomposition of TC and MB with time. Removal of TC when carried out in batch (A) and semi-batch (B) setup. Removal of MB when carried out in batch (C) and semi-batch (D) setup. The concentration of rough particles, MB, TC, and H${_2}$O${_2}$ employed are 100 ppm, 15 $\mu$M, 15 $\mu$M, and 5\%, respectively.}
\label{type}
\end{figure}

Next, we discuss the performance of rough particles in batch and semi-batch conditions. This study highlights the engineering processes that can substantially impact when transitioning between real-time applications. Figure \ref{type} refers to the decomposition reaction catalyzed by rough particles (Pt/PS) in the presence of a moderate concentration of 5\% H${_2}$O${_2}$. Note: The absorbance maxima at 358 nm from each UV-Vis spectral plot were utilised to determine the concentration as a function of time. The concentration of particles and the pollutants used for the comparison study are 100 ppm and 15 $\mu$M, respectively. According to the data, the highest TC elimination may be accomplished when the reaction is conducted under semi-batch conditions. On the other hand, when similar decomposition kinetics was run under batch circumstances, the catalyst-assisted heterogeneous response produced a high reaction rate for MB removal, i.e., 100\% elimination of MB in 10 minutes. These results intrigued us to set up reaction conditions appropriately. We, therefore, systematically conducted the heterogeneous catalyst-assisted reactions using the rough particles of desired sizes and concentrations.

\begin{figure}
\centering
\includegraphics[width=\linewidth]{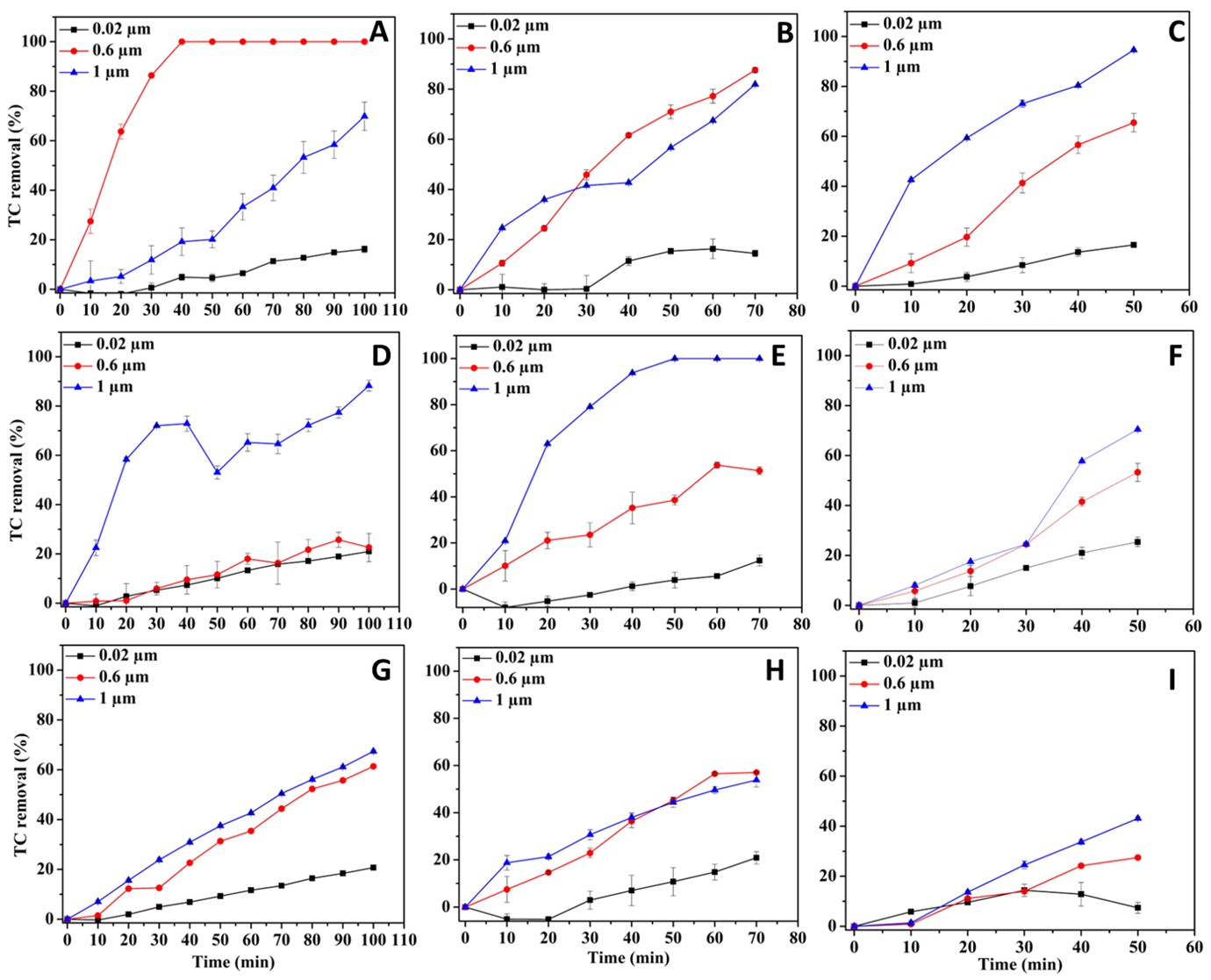}
\caption{Effect of dosing volume of H${_2}$O${_2}$ on the decomposition of TC with time under semi-batch mode at regular interval. A-C) Removal of TC at different dosing volume of 0.416, 0.624, and 0.833 mL of H${_2}$O${_2}$, respectively. D-F) Removal of TC at different dosing volume of 0.833, 1.25, and 1.66 mL of H${_2}$O${_2}$, respectively. G-I) Removal of TC at different dosing volume of 1.66, 2.5, and 3.33 mL of H${_2}$O${_2}$, respectively. The concentration of rough particles used are 200 ppm (A-C), 100 ppm (D-F), and 50 ppm (G-I), respectively. The initial concentration of TC employed for the study is 15 $\mu$M, throughout.}
\label{TC}
\end{figure}

To this end, we demonstrate the performance of rough particles in semi-batch conditions to determine the optimum parameters. Since the semi-batch operation was effective for TC, we conducted all experiments related to TC removal using a semi-batch set-up to find the optimal point. Figure \ref{TC} displays the kinetics of the decomposition reaction of TC carried out in semi-batch operation. By comparing Figures \ref{type} and \ref{TC}, the kinetic data suggest that pollutants' concentration has an inverse effect on the reaction rate associated with TC removal. Therefore, the semi-batch catalyzed reaction favours maximal elimination at a higher rate. In contrast to batch mode, this condition improved the reaction rate because TC concentrations were kept as low as possible due to drop-wise addition. We infer from Figure \ref{TC} that the maximum efficiency is directly proportional to particle concentration. Table \ref{Efficiency_TC} summarizes the performance of rough particles employed to deactivate TC. As shown in Figure \ref{TC} and Table \ref{Efficiency_TC}, the rough particles with a size of 0.6 $\mu$m and concentration 200 ppm along with a dosing volume of 0.416 mL H${_2}$O${_2}$ exhibited the best performance among the given set of rough particles and the experimental conditions.

\begin{table}[htbp]
  \centering
   \scalebox{0.7}{
  \caption{Summary of performances of the rough particles over removal of TC}
  \label{Efficiency_TC}%
    \begin{tabular}{ccccccc}
    \toprule
    Type  & Size ($\mu$m)  & Initial TC conc. ($\mu$M) & Particles conc. (ppm) & Dosing volume (mL) & Max. Efficiency (\%) & Time (min) \\
    \midrule
    \multirow{3}[2]{*}{Pt-PS} & 1     & \multirow{3}[2]{*}{15} & 100   & 1.25 & 100   & 50 \\
          & 0.6   &       & 200   & 0.416  & 100   & 40 \\
          & 0.02  &       & 40    & 1.66  & 65    & 100 \\
    \bottomrule
    \end{tabular}}%
 \end{table}%

\begin{figure}
\centering
\includegraphics[width=\linewidth]{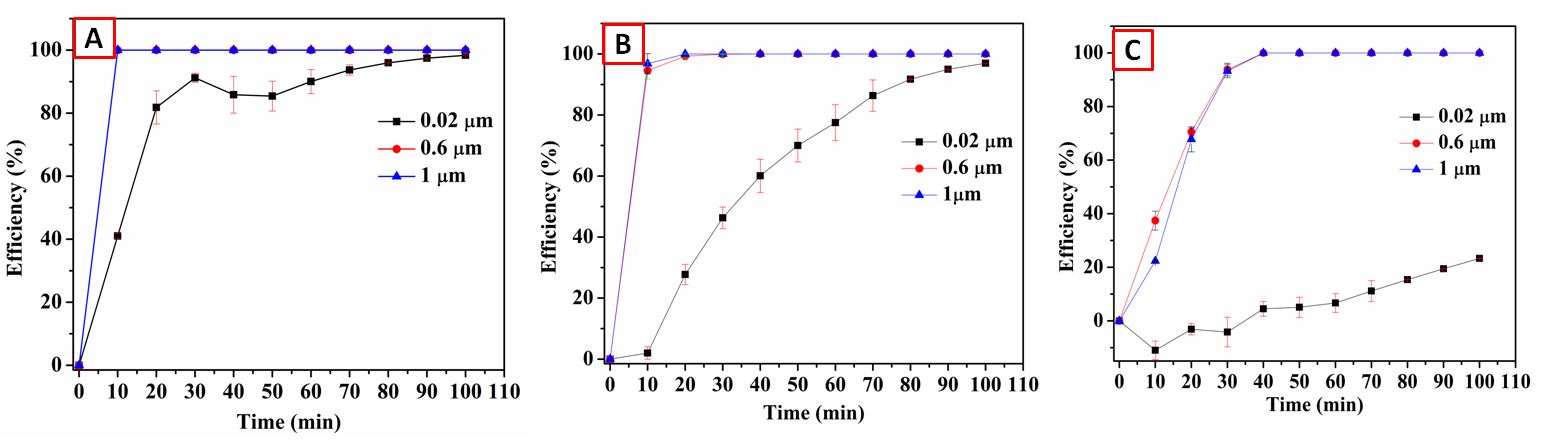}
\caption{Removal of MB catalyzed by varying sizes of rough particles at a concentration of A) 200 ppm, B) 100 ppm, and C) 50 ppm.}
\label{MB}
\end{figure}

Further, we discuss the performance of rough particles in batch conditions. Since the batch operation was effective for MB, we ran all experiments related to MB removal using a batch mode to determine the optimal point. Figure \ref{MB} displays the kinetics of the decomposition reaction of MB carried out in batch operation. Note: The absorbance maxima at 664 nm from each UV-Vis spectral plot were utilized to determine the concentration as a function of time. We infer that the performances of the rough particles of size 1.0 and 0.6 $\mu$m, respectively, overlap significantly over time. Barring 0.02 $\mu$m, the rate of decomposition data shows direct dependency on the concentration of rough particles. Therefore, the batch-catalyzed reaction favours the maximal elimination rate at a higher concentration of rough particles. In contrast to the semi-batch mode, this condition improved the reaction rate because the concentration of MB was kept high at the start. Table \ref{Efficiency_MB} summarizes the performance of rough particles employed to deactivate MB. As shown in Figure \ref{MB} and Table \ref{Efficiency_MB}, the rough particles with a size of 1 and 0.6 $\mu$m and concentration 200 ppm at a concentration of 5\% H${_2}$O${_2}$ exhibited the best performance among the given set of rough particles and the experimental conditions.

\begin{table}[htbp]
  \centering
   \scalebox{0.85}{
  \caption{Summary of performances of rough particles against MB removal at 5\% H${_2}$O${_2}$}
  \label{Efficiency_MB}
   \begin{tabular}{cccccc}
    \toprule
    Type  & Size ($\mu$m)  & Initial MB conc. ($\mu$M) & Particle conc. (ppm) & Max. Efficiency (\%) & Time (min) \\
    \midrule
    \multirow{3}[2]{*}{PS} & \multirow{3}[2]{*}{1 \& 0.6} & \multirow{3}[2]{*}{15} & 200   & 100   & 10 \\
          &       &       & 100   & 100   & 20 \\
          &       &       & 50    & 100   & 40 \\
    \bottomrule
    \end{tabular}}%
  \end{table}%

Besides \% of elimination, the order of the reaction and rate constant also provide good insight into the kinetics of the Pt-PS-assisted reactions. Langmuir-Hinshelwood (LH) kinetics is the most popular model for describing the kinetics of heterogeneous catalytic processes which is described as shown in Eq. \ref{Eq1} below\cite{KUMAR200882}, 

\begin{equation}
    r=-\frac{dC}{dt}=\frac{k_rKC}{1+KC}
    \label{Eq1}
\end{equation}

where r, k$_{r}$, K, and C refer to the rate of reaction that changes with time, limiting rate constant of reaction at maximum coverage under the given experimental conditions, equilibrium constant for adsorption of the substrate onto the catalyst, and concentration at anytime t during degradation, respectively.

\begin{figure}
\centering
\includegraphics[width=\linewidth]{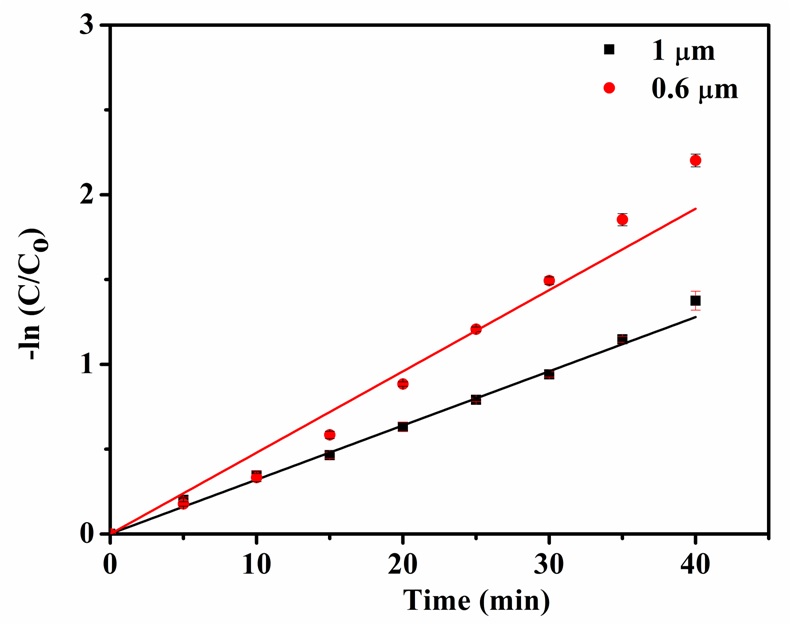}
\caption{Pseudo first-order kinetic plot for the deactivation of MB in the presence of rough particles of size 0.6 and 1.0 $\mu$m.}
\label{Kinetic}
\end{figure}

Most of the time, researchers approximated Eq. \ref{Eq1} to the first order, i.e., n=1, by assuming KC $<< 1$. Thus, when $-\mathbf{ln}{\left(\frac{C}{C_0}\right)}$ is plotted on the ordinate and time is plotted on the abscissa, the slope of the straight line is the product of k$_{r}$ and K. However, if the value of the slope found is $>>$ 1, then the assumption that KC $<<$ 1 becomes invalid. That means the subsequent kinetics will be in zero order. Figure \ref{Kinetic} displays the pseudo-first-order kinetics corresponding to the heterogeneous reactions catalyzed by the rough particles of varying sizes. The rate constants for the reaction catalyzed by 0.6 and 1.0 $\mu$m particles at 50 ppm are 0.048 and 0.032 min$^{-1}$, respectively. The reported values are in the same order as that of the performance of the Fe-Pt nanoparticles reported by \citeauthor{hsieh2012fept} for 5 ppm \cite{hsieh2012fept}. Due to variations in the concentration of the rough particles, the reported values in our case are 1.4 to 2.0 times higher than the literature value \cite{hsieh2012fept}. It is also noteworthy that the process catalyzed by the rough particles at a concentration of 50 ppm completes (100\%) at the end of the 40-minute reaction time. By 90 minutes, the absorbance maximum near 665 nm for the Fe-Pt particles exhibited by \citeauthor{hsieh2012fept} had dropped by just 90 \% \cite{hsieh2012fept}.

\begin{figure}
\centering
\includegraphics[width=\linewidth]{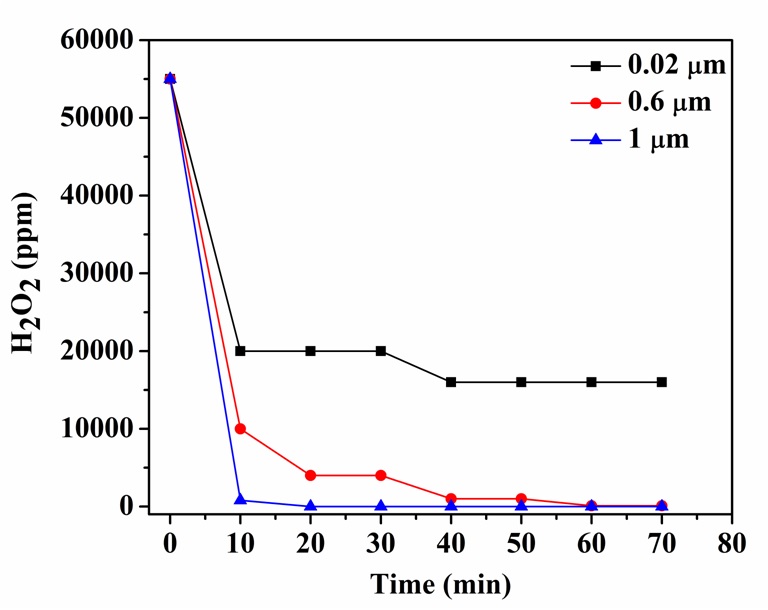}
\caption{H${_2}$O${_2}$ decomposition using the rough particles of varying sizes.}
\label{peroxide}
\end{figure}

Before moving on to the following section and exhibiting the proof-of-concept employing magnetically responsive particles, we propose probable causes for the disparity in performance between the different-sized rough particles. It is generally understood that the nanoparticles offer a more significant surface area (SA) for a given concentration in a batch reactor. For instance, we anticipate that the smaller rough particles would expose Pt significantly due to the more significant number of particles than the same concentration of bigger rough particles. By referring to Table \ref{table} for zeta potential values and EDX analysis of the rough particles of varying sizes, we cancel out any variation associated with the quality of the Pt deposition. Consequently, the SA of rough particles should increase as their size decreases, i.e. 1.0 $\mu$m > 0.6 $\mu$m > 0.02 $\mu$m. In contrast, according to Figures \ref{MB} and \ref{Kinetic}, we observed a kinetics that is an inverse function of particle size when examining the decomposition of MB using the rough particles under batch conditions. We attribute this unusual behaviour to the crowding effect of bubbles and the buoyancy effect of the nanoparticles caused due to the attachment of bubbles. As a result, 1) particles diffuse rapidly towards the interface, decreasing the contact time with a key reactant, and 2) the catalyst's power is diminished due to a layer of bubbles around the particles obstructing the active sites. Figure \ref{peroxide} displays the peroxide decomposition reaction as a function of time in minutes. Except for 0.6 and 1.0 $\mu$m, we found a marked difference in performance concerning 0.02 $\mu$m. The differences in peroxide breakdown kinetics found using the detector strips could be linked to the aforementioned factors. The video showing the rapid generation of the bubbles and the movement of the nanoparticles from bulk to the interface can be found in Supplementary Information (SI). 

\subsection{Proof-of-concept}
Thus far, we have demonstrated a straightforward methodology to make the rough particles by modifying the surface of the PS using the Pt nanoparticles through the wet-chemical route and their performances in decontamination application. Here, we show that the magnetically-responsive rough particles (MR-RP) can be made by decorating platinum-coated PS nanoparticles with iron oxide nanoparticles like a core-shell structure to accomplish chemical deactivation and particle recovery in a single step. The benefits of MR-RP include ease of fabrication of Pt nanoparticles as the deposition takes place by attractive electrostatic forces between Pt precursor (negatively charged) and the positively charged PS particles. This combination helps us achieve a uniform deposition like a monolayer of islands distributed in space throughout without needing stabilizers like oleic acid and oleyl amine. Secondly, unlike the method proposed by \citeauthor {hsieh2012fept}\cite{hsieh2012fept}, the technique does not demand heating at high temperature, i.e., 297$^{\circ}$C, and the entire process can be completed in room temperature. 

To this end, we demonstrate the discolourization of MB using MR-RP. Scheme \ref{scheme2} describes the process of making MR-RP using a wet-chemical route. Since the process follows the modifications in bulk, we can eliminate the shortcomings such as scale-up and yield. As shown in Scheme \ref{scheme2}, the attractive forces between IO and Polydiallyldimethylammonium chloride (PolyDADMAC) allow the binding of the polymers on the surface of IO nanoparticles. Subsequently, the suspension containing Pt-modified PS nanoparticles of size 0.02 $\mu$m was introduced into the polymer-coated IO mixture. This addition results in the self-assembly of negatively charged Pt-modified PS and positively charged polymer-modified IO leading to the formation of core-shell particles. Note: Since the sufficient concentration of PS nanoparticles was considered during the wet-chemical deposition and a suitable ratio of Pt-decorated PS and polymer-decorated IO was chosen before mixing, there were very little or no free Pt nanoparticles found in the supernatant solution
(Please see supplementary video-2 shown in the supplementary information). We demonstrate a start-to-finish proof-of-concept of a two-step process (removal of MB + recovery of MR-RP) achieved by the proposed surface-engineered nanoparticles via a real-time video  (Please see supplementary video-3 shown in the supplementary information).  

\begin{scheme}
\centering
\includegraphics[width=\linewidth]{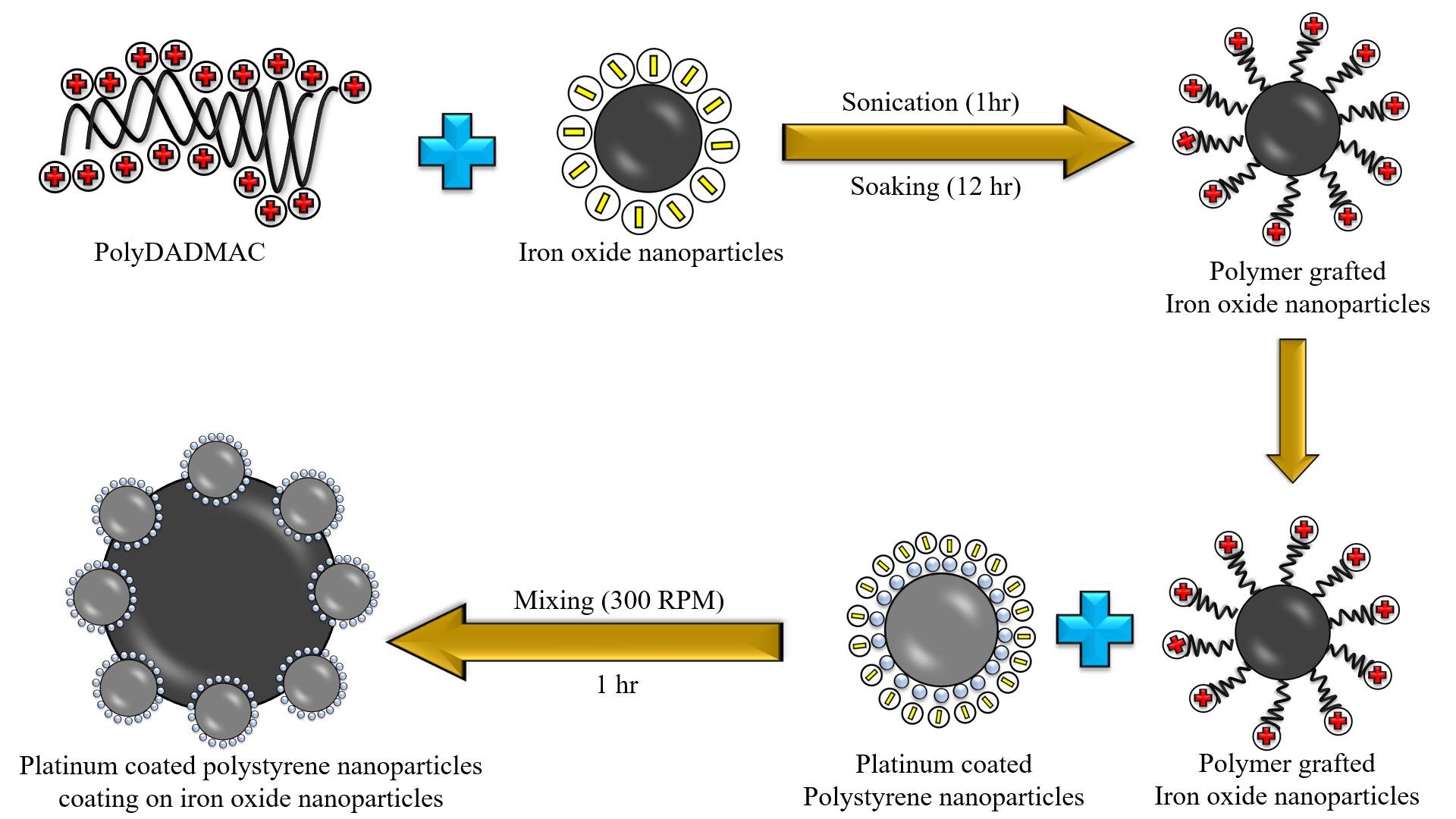}
\caption{Schematic description showing the process of synthesizing magnetically-responsive rough particles using the wet-chemical method.}
\label{scheme2}
\end{scheme}

\section{General Remarks}
We comment on the proposed method's applicability and shortcomings in this section. The proposed method works well against pollutants that accept electrons for chemical deactivation. For instance, the process would not fit well for the methyl orange dye, a contaminant that would give up electrons while decomposing. We tested the applicability of the rough particles against methyl orange and found that the decomposition rate is relatively slow compared to the reaction of its counterparts. Thus, we understand that the proposed rough particles show catalytic power and facilitate electron transfer from free radicals to pollutants over time. Secondly, the proposed method requires PS particles to be positively charged to attract the precursor ions, which are negatively charged and induce nucleation of platinum nanoparticles on the surface of PS. Therefore, the method is unsuitable for making rough particles if the terminal end groups of PS are negatively charged.

\section{Conclusion}
We established a facile approach for producing Pt-studded PS-based rough particles with good catalytic activity in a single step. The proposed method offers an exciting route to make catalyst particles on desired supports at room temperature. As the existing process requires chemical precipitation followed by heating the reaction mixture to 297$^{\circ}$C, our proposed approach could serve as an excellent alternative to existing Fenton-based heterogeneous particles studded with Pt metal nanoparticles. Our study revealed that the contacting patterns play a significant role in determining the performance of the rough particles over the decomposition of any given pollutant. For instance, the catalytic action of PS-Pt rough particles was visible when tetracycline decomposition was performed in semi-batch rather than batch settings. Conversely, batch-wise operations recorded the maximal output of rough particles in eliminating methylene blue. We studied the impact of particle sizes, concentrations and dosing rate of hydrogen peroxide on the catalytic activity of rough particles. To cite an example, the desired operating parameters to achieve 100\% efficiency vary for different rough particles of size 0.6 $\mu$m (Optimum concentration = 200 ppm, dosing volume = 0.416 mL) and 1$\mu$m (Optimum concentration = 100 ppm, dosing volume = 1.25 mL). The heterogeneous reaction modelled by Langmuir-Hinshelwood kinetics indicates that the catalytic reaction follows the pseudo-first-order process with rate constants of 0.048 and 0.032 min$^{-1}$ for the reaction catalyzed by 0.6 and 1.0 $\mu$m-sized rough particles, respectively. Using a well-known layer-by-layer technique, we presented a straightforward methodology to achieve a core-shell assembly of magnetically-responsive rough particles (MR-RP). In a batch reactor configuration, we demonstrated proof-of-concept for the decomposition of methylene blue using MR-RP. Since we can deactivate pollutants and recover catalyst particles in a single cycle, the proposed technique would be effective in real-time and large-scale applications without incurring operating expenses.  




\begin{acknowledgement}

We thank Dr Neethu Thomas (Postdoctoral Researcher) and Prof. Parasuraman Swaminathan, Department of Metallurgical and Materials Engineering, IIT Madras, India, for helping us with the FESEM combined with the EDX analysis of our samples. MS thanks IIT Ropar for providing the seed grant (ISIRD phase II) to set up a laboratory. 

\end{acknowledgement}






\providecommand{\latin}[1]{#1}
\makeatletter
\providecommand{\doi}
  {\begingroup\let\do\@makeother\dospecials
  \catcode`\{=1 \catcode`\}=2 \doi@aux}
\providecommand{\doi@aux}[1]{\endgroup\texttt{#1}}
\makeatother
\providecommand*\mcitethebibliography{\thebibliography}
\csname @ifundefined\endcsname{endmcitethebibliography}
  {\let\endmcitethebibliography\endthebibliography}{}
\begin{mcitethebibliography}{23}
\providecommand*\natexlab[1]{#1}
\providecommand*\mciteSetBstSublistMode[1]{}
\providecommand*\mciteSetBstMaxWidthForm[2]{}
\providecommand*\mciteBstWouldAddEndPuncttrue
  {\def\EndOfBibitem{\unskip.}}
\providecommand*\mciteBstWouldAddEndPunctfalse
  {\let\EndOfBibitem\relax}
\providecommand*\mciteSetBstMidEndSepPunct[3]{}
\providecommand*\mciteSetBstSublistLabelBeginEnd[3]{}
\providecommand*\EndOfBibitem{}
\mciteSetBstSublistMode{f}
\mciteSetBstMaxWidthForm{subitem}{(\alph{mcitesubitemcount})}
\mciteSetBstSublistLabelBeginEnd
  {\mcitemaxwidthsubitemform\space}
  {\relax}
  {\relax}

\bibitem[Visa and Chelaru(2014)Visa, and Chelaru]{visa2014hydrothermally}
Visa,~M.; Chelaru,~A.-M. Hydrothermally modified fly ash for heavy metals and
  dyes removal in advanced wastewater treatment. \emph{Applied Surface Science}
  \textbf{2014}, \emph{303}, 14--22\relax
\mciteBstWouldAddEndPuncttrue
\mciteSetBstMidEndSepPunct{\mcitedefaultmidpunct}
{\mcitedefaultendpunct}{\mcitedefaultseppunct}\relax
\EndOfBibitem
\bibitem[Liu \latin{et~al.}(2021)Liu, Chen, Zhang, Shan, Wu, Bai, and
  Wang]{liu2021treatment}
Liu,~L.; Chen,~Z.; Zhang,~J.; Shan,~D.; Wu,~Y.; Bai,~L.; Wang,~B. Treatment of
  industrial dye wastewater and pharmaceutical residue wastewater by advanced
  oxidation processes and its combination with nanocatalysts: A review.
  \emph{Journal of Water Process Engineering} \textbf{2021}, \emph{42},
  102122\relax
\mciteBstWouldAddEndPuncttrue
\mciteSetBstMidEndSepPunct{\mcitedefaultmidpunct}
{\mcitedefaultendpunct}{\mcitedefaultseppunct}\relax
\EndOfBibitem
\bibitem[Minero \latin{et~al.}(2005)Minero, Lucchiari, Vione, and
  Maurino]{minero2005fe}
Minero,~C.; Lucchiari,~M.; Vione,~D.; Maurino,~V. Fe (III)-enhanced
  sonochemical degradation of methylene blue in aqueous solution.
  \emph{Environmental science \& technology} \textbf{2005}, \emph{39},
  8936--8942\relax
\mciteBstWouldAddEndPuncttrue
\mciteSetBstMidEndSepPunct{\mcitedefaultmidpunct}
{\mcitedefaultendpunct}{\mcitedefaultseppunct}\relax
\EndOfBibitem
\bibitem[Peter \latin{et~al.}(2017)Peter, Mihaly-Cozmuta, Nicula,
  Mihaly-Cozmuta, Jastrz{\k{e}}bska, Olszyna, and Baia]{peter2017uv}
Peter,~A.; Mihaly-Cozmuta,~A.; Nicula,~C.; Mihaly-Cozmuta,~L.;
  Jastrz{\k{e}}bska,~A.; Olszyna,~A.; Baia,~L. UV light-assisted degradation of
  methyl orange, methylene blue, phenol, salicylic acid, and rhodamine B:
  photolysis versus photocatalyis. \emph{Water, Air, \& Soil Pollution}
  \textbf{2017}, \emph{228}, 1--12\relax
\mciteBstWouldAddEndPuncttrue
\mciteSetBstMidEndSepPunct{\mcitedefaultmidpunct}
{\mcitedefaultendpunct}{\mcitedefaultseppunct}\relax
\EndOfBibitem
\bibitem[Mohammed \latin{et~al.}(2021)Mohammed, Khaleefa, and
  Basheer]{mohammed2021photolysis}
Mohammed,~H.~A.; Khaleefa,~S.~A.; Basheer,~M.~I. Photolysis of methylene blue
  dye using an advanced oxidation process (ultraviolet light and hydrogen
  peroxide). \emph{Journal of Engineering and Sustainable Development}
  \textbf{2021}, \emph{25}\relax
\mciteBstWouldAddEndPuncttrue
\mciteSetBstMidEndSepPunct{\mcitedefaultmidpunct}
{\mcitedefaultendpunct}{\mcitedefaultseppunct}\relax
\EndOfBibitem
\bibitem[Hsieh and Lin(2012)Hsieh, and Lin]{hsieh2012fept}
Hsieh,~S.; Lin,~P.-Y. FePt nanoparticles as heterogeneous Fenton-like catalysts
  for hydrogen peroxide decomposition and the decolorization of methylene blue.
  \emph{Journal of Nanoparticle Research} \textbf{2012}, \emph{14}, 1--10\relax
\mciteBstWouldAddEndPuncttrue
\mciteSetBstMidEndSepPunct{\mcitedefaultmidpunct}
{\mcitedefaultendpunct}{\mcitedefaultseppunct}\relax
\EndOfBibitem
\bibitem[Saleh and Taufik(2019)Saleh, and Taufik]{saleh2019degradation}
Saleh,~R.; Taufik,~A. Degradation of methylene blue and congo-red dyes using
  Fenton, photo-Fenton, sono-Fenton, and sonophoto-Fenton methods in the
  presence of iron (II, III) oxide/zinc oxide/graphene (Fe3O4/ZnO/graphene)
  composites. \emph{Separation and Purification Technology} \textbf{2019},
  \emph{210}, 563--573\relax
\mciteBstWouldAddEndPuncttrue
\mciteSetBstMidEndSepPunct{\mcitedefaultmidpunct}
{\mcitedefaultendpunct}{\mcitedefaultseppunct}\relax
\EndOfBibitem
\bibitem[Kirchon \latin{et~al.}(2020)Kirchon, Zhang, Li, Joseph, Chen, and
  Zhou]{kirchon2020effect}
Kirchon,~A.; Zhang,~P.; Li,~J.; Joseph,~E.~A.; Chen,~W.; Zhou,~H.-C. Effect of
  isomorphic metal substitution on the fenton and photo-fenton degradation of
  methylene blue using Fe-based metal--organic frameworks. \emph{ACS applied
  materials \& interfaces} \textbf{2020}, \emph{12}, 9292--9299\relax
\mciteBstWouldAddEndPuncttrue
\mciteSetBstMidEndSepPunct{\mcitedefaultmidpunct}
{\mcitedefaultendpunct}{\mcitedefaultseppunct}\relax
\EndOfBibitem
\bibitem[Ikhlaqa \latin{et~al.}(2020)Ikhlaqa, Javedb, Niaza, Munirc, and
  Qid]{ikhlaqa2020combined}
Ikhlaqa,~A.; Javedb,~F.; Niaza,~A.; Munirc,~H. M.~S.; Qid,~F. Combined UV
  catalytic ozonation process on iron loaded peanut shell ash for the removal
  of methylene blue from aqueous solution. \emph{DESALINATION AND WATER
  TREATMENT} \textbf{2020}, \emph{200}, 231--240\relax
\mciteBstWouldAddEndPuncttrue
\mciteSetBstMidEndSepPunct{\mcitedefaultmidpunct}
{\mcitedefaultendpunct}{\mcitedefaultseppunct}\relax
\EndOfBibitem
\bibitem[Adelin \latin{et~al.}(2020)Adelin, Gunawan, Nur, Haris, Widodo, and
  Suyati]{adelin2020ozonation}
Adelin,~M.; Gunawan,~G.; Nur,~M.; Haris,~A.; Widodo,~D.; Suyati,~L. Ozonation
  of methylene blue and its fate study using LC-MS/MS. Journal of Physics:
  Conference Series. 2020; p 012079\relax
\mciteBstWouldAddEndPuncttrue
\mciteSetBstMidEndSepPunct{\mcitedefaultmidpunct}
{\mcitedefaultendpunct}{\mcitedefaultseppunct}\relax
\EndOfBibitem
\bibitem[Qian \latin{et~al.}(2019)Qian, Blatov, Wang, Ding, Zhu, Li, Li, and
  Wu]{qian2019sonochemical}
Qian,~L.-L.; Blatov,~V.~A.; Wang,~Z.-X.; Ding,~J.-G.; Zhu,~L.-M.; Li,~K.;
  Li,~B.-L.; Wu,~B. Sonochemical synthesis and characterization of four
  nanostructural nickel coordination polymers and photocatalytic degradation of
  methylene blue. \emph{Ultrasonics sonochemistry} \textbf{2019}, \emph{56},
  213--228\relax
\mciteBstWouldAddEndPuncttrue
\mciteSetBstMidEndSepPunct{\mcitedefaultmidpunct}
{\mcitedefaultendpunct}{\mcitedefaultseppunct}\relax
\EndOfBibitem
\bibitem[Liu \latin{et~al.}(2016)Liu, He, Fu, and
  Dionysiou]{liu2016degradation}
Liu,~Y.; He,~X.; Fu,~Y.; Dionysiou,~D.~D. Degradation kinetics and mechanism of
  oxytetracycline by hydroxyl radical-based advanced oxidation processes.
  \emph{Chemical Engineering Journal} \textbf{2016}, \emph{284},
  1317--1327\relax
\mciteBstWouldAddEndPuncttrue
\mciteSetBstMidEndSepPunct{\mcitedefaultmidpunct}
{\mcitedefaultendpunct}{\mcitedefaultseppunct}\relax
\EndOfBibitem
\bibitem[Pignatello \latin{et~al.}(2006)Pignatello, Oliveros, and
  MacKay]{pignatello2006advanced}
Pignatello,~J.~J.; Oliveros,~E.; MacKay,~A. Advanced oxidation processes for
  organic contaminant destruction based on the Fenton reaction and related
  chemistry. \emph{Critical reviews in environmental science and technology}
  \textbf{2006}, \emph{36}, 1--84\relax
\mciteBstWouldAddEndPuncttrue
\mciteSetBstMidEndSepPunct{\mcitedefaultmidpunct}
{\mcitedefaultendpunct}{\mcitedefaultseppunct}\relax
\EndOfBibitem
\bibitem[Giordano \latin{et~al.}(2007)Giordano, Perathoner, Centi, {De Rosa},
  Granato, Katovic, Siciliano, Tagarelli, and Tripicchio]{GIORDANO2007240}
Giordano,~G.; Perathoner,~S.; Centi,~G.; {De Rosa},~S.; Granato,~T.;
  Katovic,~A.; Siciliano,~A.; Tagarelli,~A.; Tripicchio,~F. Wet hydrogen
  peroxide catalytic oxidation of olive oil mill wastewaters using Cu-zeolite
  and Cu-pillared clay catalysts. \emph{Catalysis Today} \textbf{2007},
  \emph{124}, 240--246, Advanced Catalytic Oxidation Processes\relax
\mciteBstWouldAddEndPuncttrue
\mciteSetBstMidEndSepPunct{\mcitedefaultmidpunct}
{\mcitedefaultendpunct}{\mcitedefaultseppunct}\relax
\EndOfBibitem
\bibitem[Ai \latin{et~al.}(2007)Ai, Lu, Li, Zhang, Qiu, and Wu]{ai2007fe}
Ai,~Z.; Lu,~L.; Li,~J.; Zhang,~L.; Qiu,~J.; Wu,~M. Fe@ Fe2O3 core- shell
  nanowires as iron reagent. 1. Efficient degradation of rhodamine B by a novel
  sono-Fenton process. \emph{The Journal of Physical Chemistry C}
  \textbf{2007}, \emph{111}, 4087--4093\relax
\mciteBstWouldAddEndPuncttrue
\mciteSetBstMidEndSepPunct{\mcitedefaultmidpunct}
{\mcitedefaultendpunct}{\mcitedefaultseppunct}\relax
\EndOfBibitem
\bibitem[Toda \latin{et~al.}(1999)Toda, Igarashi, Uchida, and
  Watanabe]{Toda_1999}
Toda,~T.; Igarashi,~H.; Uchida,~H.; Watanabe,~M. Enhancement of the
  Electroreduction of Oxygen on Pt Alloys with Fe, Ni, and Co. \emph{Journal of
  The Electrochemical Society} \textbf{1999}, \emph{146}, 3750--3756\relax
\mciteBstWouldAddEndPuncttrue
\mciteSetBstMidEndSepPunct{\mcitedefaultmidpunct}
{\mcitedefaultendpunct}{\mcitedefaultseppunct}\relax
\EndOfBibitem
\bibitem[Dokoutchaev \latin{et~al.}(1999)Dokoutchaev, James, Koene, Pathak,
  Prakash, and Thompson]{dokoutchaev1999colloidal}
Dokoutchaev,~A.; James,~J.~T.; Koene,~S.~C.; Pathak,~S.; Prakash,~G.~S.;
  Thompson,~M.~E. Colloidal metal deposition onto functionalized polystyrene
  microspheres. \emph{Chemistry of Materials} \textbf{1999}, \emph{11},
  2389--2399\relax
\mciteBstWouldAddEndPuncttrue
\mciteSetBstMidEndSepPunct{\mcitedefaultmidpunct}
{\mcitedefaultendpunct}{\mcitedefaultseppunct}\relax
\EndOfBibitem
\bibitem[Lee \latin{et~al.}(2009)Lee, Mahmoud, Sitterle, Sitterle, and
  Meredith]{lee2009facile}
Lee,~J.-H.; Mahmoud,~M.~A.; Sitterle,~V.; Sitterle,~J.; Meredith,~J.~C. Facile
  preparation of highly-scattering metal nanoparticle-coated polymer microbeads
  and their surface plasmon resonance. \emph{Journal of the American Chemical
  Society} \textbf{2009}, \emph{131}, 5048--5049\relax
\mciteBstWouldAddEndPuncttrue
\mciteSetBstMidEndSepPunct{\mcitedefaultmidpunct}
{\mcitedefaultendpunct}{\mcitedefaultseppunct}\relax
\EndOfBibitem
\bibitem[Bao \latin{et~al.}(2014)Bao, Bihr, Smith, and Taylor]{bao2014facile}
Bao,~H.; Bihr,~T.; Smith,~A.-S.; Taylor,~R. N.~K. Facile colloidal coating of
  polystyrene nanospheres with tunable gold dendritic patches. \emph{Nanoscale}
  \textbf{2014}, \emph{6}, 3954--3966\relax
\mciteBstWouldAddEndPuncttrue
\mciteSetBstMidEndSepPunct{\mcitedefaultmidpunct}
{\mcitedefaultendpunct}{\mcitedefaultseppunct}\relax
\EndOfBibitem
\bibitem[Sabapathy \latin{et~al.}(2015)Sabapathy, Kollabattula, Basavaraj, and
  Mani]{sabapathy2015visualization}
Sabapathy,~M.; Kollabattula,~V.; Basavaraj,~M.~G.; Mani,~E. Visualization of
  the equilibrium position of colloidal particles at fluid--water interfaces by
  deposition of nanoparticles. \emph{Nanoscale} \textbf{2015}, \emph{7},
  13868--13876\relax
\mciteBstWouldAddEndPuncttrue
\mciteSetBstMidEndSepPunct{\mcitedefaultmidpunct}
{\mcitedefaultendpunct}{\mcitedefaultseppunct}\relax
\EndOfBibitem
\bibitem[Lu \latin{et~al.}(2003)Lu, Sun, Xi, Wang, Zhang, Wang, and
  Zhou]{lu2003colloidal}
Lu,~L.; Sun,~G.; Xi,~S.; Wang,~H.; Zhang,~H.; Wang,~T.; Zhou,~X. A colloidal
  templating method to hollow bimetallic nanostructures. \emph{Langmuir}
  \textbf{2003}, \emph{19}, 3074--3077\relax
\mciteBstWouldAddEndPuncttrue
\mciteSetBstMidEndSepPunct{\mcitedefaultmidpunct}
{\mcitedefaultendpunct}{\mcitedefaultseppunct}\relax
\EndOfBibitem
\bibitem[Kumar \latin{et~al.}(2008)Kumar, Porkodi, and Rocha]{KUMAR200882}
Kumar,~K.~V.; Porkodi,~K.; Rocha,~F. Langmuir–Hinshelwood kinetics – A
  theoretical study. \emph{Catalysis Communications} \textbf{2008}, \emph{9},
  82--84\relax
\mciteBstWouldAddEndPuncttrue
\mciteSetBstMidEndSepPunct{\mcitedefaultmidpunct}
{\mcitedefaultendpunct}{\mcitedefaultseppunct}\relax
\EndOfBibitem
\end{mcitethebibliography}
\providecommand{\latin}[1]{#1}
\makeatletter
\providecommand{\doi}
  {\begingroup\let\do\@makeother\dospecials
  \catcode`\{=1 \catcode`\}=2 \doi@aux}
\providecommand{\doi@aux}[1]{\endgroup\texttt{#1}}
\makeatother
\providecommand*\mcitethebibliography{\thebibliography}
\csname @ifundefined\endcsname{endmcitethebibliography}
  {\let\endmcitethebibliography\endthebibliography}{}

\end{sloppypar}

\end{document}